\documentclass[10pt,conference]{IEEEtran}
  \IEEEoverridecommandlockouts
  
\usepackage{cite}
\usepackage[symbol]{footmisc}

\usepackage{graphicx}
{\fnsymbol{footnote}}
\usepackage{amsmath,amssymb,amsfonts}
\usepackage{algorithmic}
\usepackage{graphicx}
\usepackage{textcomp}
\usepackage{xcolor}
\IEEEoverridecommandlockouts
 \usepackage {cite}
 \usepackage{amsmath, amssymb} 
\usepackage[linesnumbered,ruled,vlined]{algorithm2e}
\usepackage{amsmath}
\usepackage{graphicx}
\usepackage{epstopdf}
\usepackage{setspace}
\usepackage{graphicx}
\usepackage{amssymb}
\usepackage{multicol}
\usepackage{amsmath}
\usepackage{color}
\usepackage{amsmath}
\usepackage{romannum}
\usepackage{epstopdf}
\usepackage{graphicx}
\usepackage{tabularx}
\usepackage{booktabs}
\usepackage{graphicx}
\usepackage{caption}
\usepackage{subcaption}
\usepackage[left=0.625in,right=0.625in,top=0.75in,bottom=1in]{geometry}
\usepackage[linesnumbered,ruled,vlined]{algorithm2e}
\newcommand*{\rom}[1]{\expandafter\@slowromancap\romannumeral #1@}
\def\BibTeX{{\rm B\kern-.05em{\sc i\kern-.025em b}\kern-.08em
    T\kern-.1667em\lower.7ex\hbox{E}\kern-.125emX}}
\begin{document}

\title{Interference Resilient Quantum Receivers with Rydberg Atoms}


\author{
\IEEEauthorblockN{Javane Rostampoor\IEEEauthorrefmark{1} and Raviraj Adve\IEEEauthorrefmark{2}}  
\IEEEauthorblockA{
\textit{Dept. of Electrical and Computer Engineering},  
\textit{University of Toronto},  
Toronto, Canada \\
Email: \IEEEauthorrefmark{1}javane.rostampoor@utoronto.ca, \IEEEauthorrefmark{2}rsadve@ece.utoronto.ca
}
}

\maketitle
\begin{abstract}
Quantum sensing has attracted significant attention due to its ability to measure physical quantities with extremely high accuracy. Rydberg atoms—typically alkali atoms with a highly excited valence electron that is far from the nucleus—exhibit strong sensitivity to external electromagnetic fields. This sensitivity leads to coupling between different atomic energy levels, which can be observed by monitoring changes in a control laser beam before and after it passes through a vapor cell containing the Rydberg atoms. By analyzing the transmitted laser signal with a photodetector, variations in transmission can be attributed to the presence and characteristics of the external electromagnetic field.
Because Rydberg atoms operate in a highly excited quantum state without relying on traditional electronic circuitry, they inherently avoid thermal noise, thereby enabling more sensitive detection. In this paper, we investigate the performance of a Rydberg atomic receiver based on \textsuperscript{85}Rb and compare it with that of a conventional receiver in detecting an 8-level pulse amplitude modulation (8-PAM) signal in the presence of off-resonant interference. We demonstrate that the Rydberg receiver can suppress interference without the need for an additional filter. Effectively, our results show that the Rydberg receiver serves as an integrated filter and demodulator, outperforming conventional circuit-based receivers in terms of achievable symbol error rate.

\end{abstract}

\begin{IEEEkeywords}
Rydberg atomic receivers, Rubidium-85, Interference, Hamiltonian, Quantum shot noise
\end{IEEEkeywords}
\section{Introduction}
Quantum communication and sensing, which are based on quantum phenomena for high-precision measurement of physical quantities, are key aspects of the second quantum revolution~\cite{EIT}. These technologies are pivotal in detecting ultra-weak biomedical signals, enabling space communication, advancing high-frequency transmissions, and more. Rydberg atoms are particularly well suited for these applications.

Rydberg atoms are atoms in a highly excited state, having a valence electron far from the nucleus and a large principal quantum number. This leads to a large electric dipole moment, making them highly sensitive to even weak electric fields and ideal for detecting subtle signals~\cite{noise}. To excite such atoms, alkali elements like Rubidium and Cesium are commonly used~\cite{alkai}. Atom-based alkali receivers for radio frequency (RF) signal detection have been widely studied~\cite{eq1source,linblad2,noise,noise2}.

Of particular relevance to our system, a resonant electromagnetic wave causes oscillation of the population between two quantum states. The frequency of this coherent population oscillation is called the \textit{Rabi frequency}, which is proportional to the amplitude of the field and the electric dipole moment of the states~\cite{Rabi}. Discovering this population fluctuation allows us to extract key information about the electromagnetic field intensity.
This mechanism underlies quantum phenomena such as Electromagnetically Induced Transparency (EIT) and Autler–Townes splitting (ATS). EIT is a quantum interference effect in a three-level atomic system, where two excitation pathways destructively interfere, suppressing absorption at a specific frequency and creating a narrow transparency window in the transmission spectrum. In Rydberg-EIT, a weak probe laser couples the ground and intermediate states, while a strong coupling laser connects the intermediate state to a highly excited Rydberg state~\cite{EIT}.


ATS happens when an RF field is applied and the Rydberg-EIT peak splits into two. This is due to the fact that any radiating electromagnetic field can alter the susceptibility of the atomic vapor as seen by the probe laser, thereby changing the probe's laser propagation. In other words, the RF signal leads to the formation of two dressed states in which the energy difference is proportional to the Rabi frequency.
Measuring this peak-to-peak separation in the transmission spectrum allows one to infer the Rabi frequency and thus the strength of the applied electromagnetic field~\cite{decay2}.

Rydberg atoms, when properly tuned, exhibit distinct transmission spectra enabling high-sensitivity RF signal detection. This is because Rydberg sensors/receivers do not require RF circuits generating thermal noise; instead, they are limited by quantum shot noise, about 15~dB lower than room temperature thermal noise~\cite{towards}. Transmission can be measured via the peak-to-peak ATS distance (frequency regime) or the transmission at resonance (amplitude regime), the latter of which, due to specific design considerations, has been less studied~\cite{Aregime}.

Another benefit of the Rydberg receiver, beyond immunity to thermal noise, is its ability to implement communication techniques as built-in features. These include amplitude and frequency demodulation by processing the probe transmission signal, and phase demodulation using a local oscillator to generate a reference field with known phase, all without requiring conventional demodulation circuitry~\cite{AMFM,Mixer}.

The primary research on atomic receivers has focused on laboratory experiments aimed at exploiting quantum features, while the signal engineering perspective remains in its infancy. The purpose of this paper is to push the boundaries of quantum receivers from a signal processing perspective. As effective interference suppression is essential for reliable signal detection, we investigate whether a Rydberg receiver can also function as an integrated filter and demodulator. 
 To the best of our knowledge, the potential of Rydberg receivers to operate as compact demodulators and filters capable of mitigating interference has not been previously studied.

In our setup, we employ a five-level system with an ${}^{85}\mathrm{Rb}$ (Rubidium-85) vapor cell to tune EIT using a probe and coupling laser, while the atoms are exposed to both an RF signal and 5G mid-band interference. Our Rydberg state preparation ensures the RF signal resonantly couples two Rydberg states, while the interference remains off-resonant. This receiver is designed to detect an 8-PAM signal in the presence of interference.
We follow the amplitude regime to read the transmission signal, which, although less studied due to its delicate design, is inherently faster and avoids the need to scan the entire spectrum. We show that our Rydberg receiver not only acts as a built-in filter, but also outperforms conventional receivers with sharp filters in terms of symbol error rate (SER).

The rest of this paper is organized as follows: the system model for a Rydberg receiver is introduced in Section~\ref{system_model}, and the noise model is discussed in Section~\ref{noise}. 
Simulation results are presented in Section~\ref{simulation} to validate the effectiveness of our proposed Rydberg receiver in cancelling interference. 
Finally, the paper is wrapped up in Section~\ref{conclusion}.

\begin{figure*}[t]  
  \centering
  \begin{subfigure}[t]{0.4\textwidth}
    \centering
    \includegraphics[width=\textwidth]{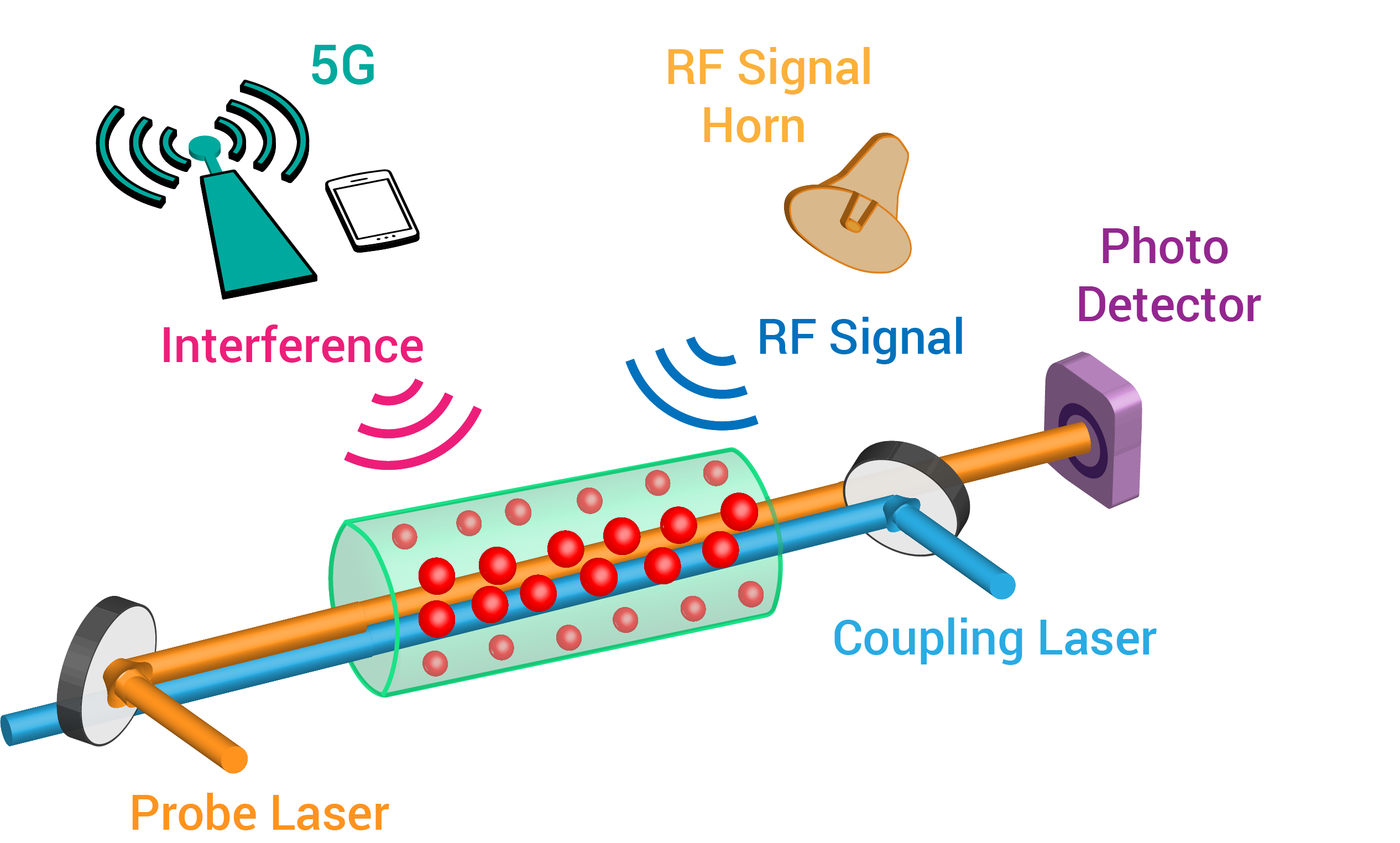}
    \caption{}
    \label{System}
  \end{subfigure}
  \hfill
  \begin{subfigure}[t]{0.34\textwidth}
    \centering
    \includegraphics[width=\textwidth]{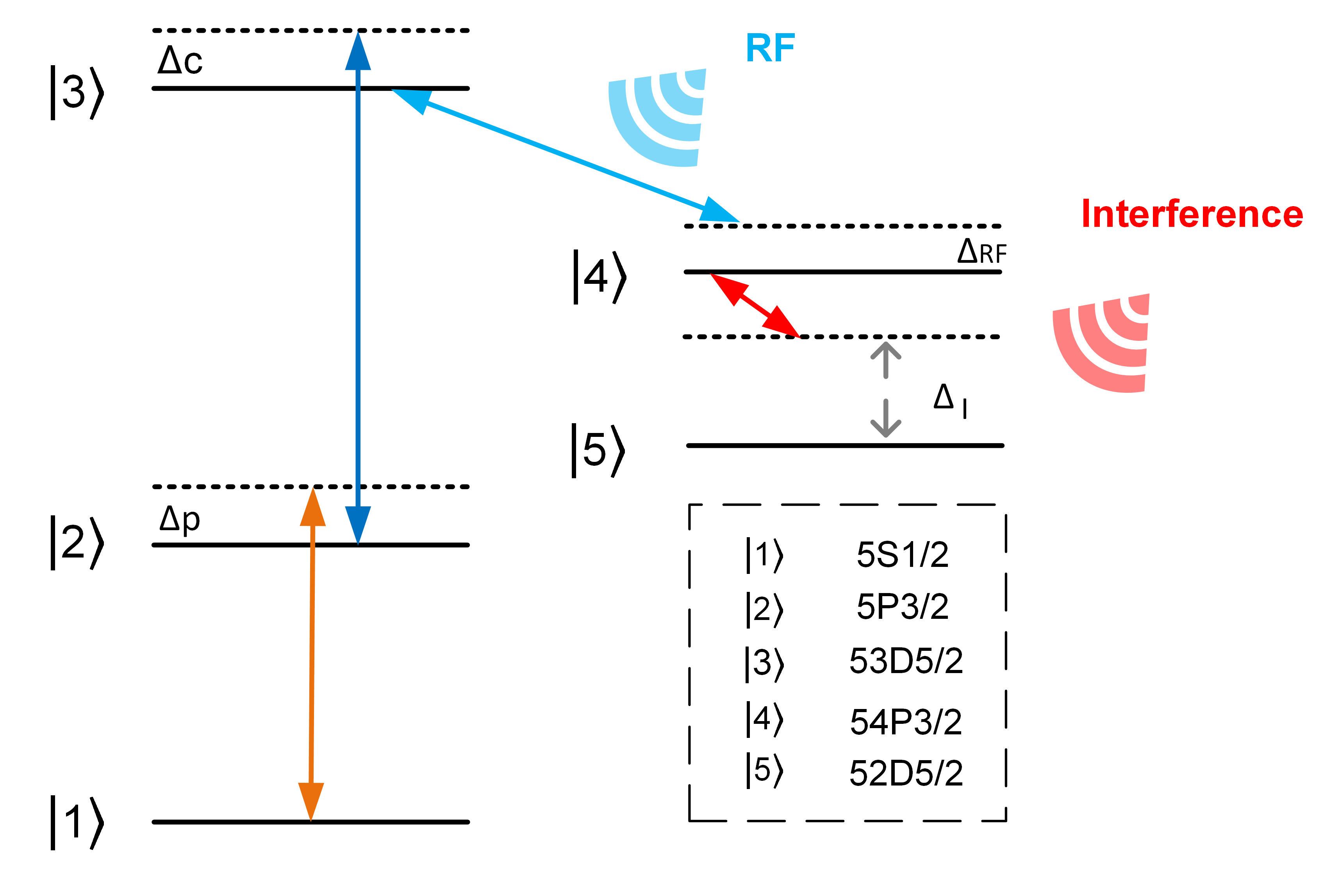}
    \caption{}
    \label{System2}
  \end{subfigure}
  \caption{(a) The schematic of the Rydberg receiver setup for detecting RF signal in the presence of 5G interference. (b) Energy level diagram of ${}^{85}\mathrm{Rb}$ with ATS in the presence of RF signal and interference.}
  \label{fig:combined}
\end{figure*}

\section{System Model}
\label{system_model}

As shown in Fig.~\ref{fig:combined}, consider an ${}^{85}\mathrm{Rb}$ vapor cell as our receiver, tuned to detect a modulated signal at $14.2\,\mathrm{GHz}$ in an environment where mid-band 5G signals (ranging from $3\,\mathrm{GHz}$ to $4\,\mathrm{GHz}$) introduce significant interference. The objective is to detect and demodulate an ultra-low-power PAM signal and simultaneously cancel interference using an all-in-one Rydberg receiver. For comparison, we evaluate the performance against a conventional receiver equipped with a filter and demodulation circuit. 

Fig.~\ref{System} shows the schematic of the setup for a Rydberg receiver, where the probe and coupling lasers are tuned to generate EIT, enabling ATS in the presence of an RF signal. Fig.~\ref{System2} illustrates the energy level couplings induced by the probe and coupling lasers, the RF signal, and the interference, outlining the structure of the Rydberg receiver. A room-temperature vapor cell containing ${}^{85}\mathrm{Rb}$ is used, in which the valence electron resides in the $5\mathrm{S}_{1/2}$ ground state. A two-photon transition is driven using a probe laser at $780\,\mathrm{nm}$ and a coupling laser at $480\,\mathrm{nm}$, which couples the intermediate state \(5\mathrm{P}_{3/2}\) to the Rydberg state \(53\mathrm{D}_{5/2}\).

The PAM-modulated RF signal resonantly drives the transition between states $\lvert 3 \rangle$ ($53\mathrm{D}_{5/2}$) and $\lvert 4 \rangle$ ($54\mathrm{P}_{3/2}$), with an energy difference corresponding to $14.2\,\mathrm{GHz}$.
 The interference lies in the mid-band 5G frequency range and is off-resonant for any nearby energy levels. Among the possible transitions, we selected the one with the highest electric dipole moment, which, in this case, is the $52\mathrm{D}_{5/2}$ level. It should be noted that, in general, the interference can also couple states $\lvert 3 \rangle$ and $\lvert 5 \rangle$; however, in this setup the dipole selection rule ($\Delta l = \pm 1$) is not satisfied.

\begin{figure*}[t]
  \centering
  \footnotesize
  \setlength{\arraycolsep}{2pt}
\begin{equation}
M =
\begin{pmatrix}
0 & \Omega_p & 0 & 0 & 0\\
\Omega_p & -2\Delta_p & \Omega_c & 0 & 0\\
0 & \Omega_c & -2(\Delta_p+\Delta_c) & \Omega_{\mathrm{RF}} & 0\\
0 & 0 & \Omega_{\mathrm{RF}} & -2(\Delta_p+\Delta_c+\Delta_{\mathrm{RF}}) & \Omega_{I}\\
0 & 0 & 0 & \Omega_{I} & -2(\Delta_p+\Delta_c+\Delta_{\mathrm{RF}}+\Delta_{I})
\end{pmatrix}
\label{M}
\end{equation}
  \label{fig:rho21}
\end{figure*}

Denoting \(\hbar\) as the reduced Planck constant and \(\Omega_i\) the Rabi frequency of each external field, and using the rotating wave approximation, the Hamiltonian is given by \(\boldsymbol{H} = \frac{\hbar}{2} \boldsymbol{M}\), where \(\boldsymbol{M}\) for a five-level system is defined in~\eqref{M} at the top of the next page~\cite{eq1source}. In this matrix, \(\Omega_i = \mu_i \bar{E}_i / \hbar\) and \(\Delta_i = \omega_i - \omega_i^*\), where \(\mu_i\) is the dipole matrix element, \(\bar{E}_i\) is the electric field amplitude, \(\omega_i\) is the angular frequency of the applied field, and \(\omega_i^*\) is the resonant angular frequency of the corresponding atomic transition. The subscript \(i\) can be \(p\), \(c\), \(\mathrm{RF}\), or \(I\), denoting the probe laser, coupling laser, RF signal, and interference signal, respectively.

The temporal evolution of the atomic density matrix $\boldsymbol{\rho}$ can be described using the Lindblad master equation~\cite{eq1source, linblad2}:
\begin{equation}
\frac{d\boldsymbol{\rho}}{dt}
= -\frac{i}{\hbar}[\boldsymbol{H},\boldsymbol{\rho}] \;+\; \mathcal{L}(\boldsymbol{\rho}),
\label{lind}
\end{equation}
where $\mathcal{L}(\boldsymbol{\rho})$ is the decay matrix, which is defined in~\eqref{L} near the top of the next page. Here, $\Gamma_i$ represents the decay rate of level $i$, and $\gamma_{i,j} = (\Gamma_i + \Gamma_j)/2$ is the decoherence rate between levels $i$ and $j$. The commutator $[\,\boldsymbol{H}, \boldsymbol{\rho}\,]$ denotes the matrix operation $\boldsymbol{H\rho} - \boldsymbol{\rho H}$.
\begin{figure*}[t]
  \centering
  \footnotesize
  \setlength{\arraycolsep}{2pt}
\begin{equation}
\mathcal{L}(\boldsymbol{\rho})  =
\begin{pmatrix}
\Gamma_2\,\rho_{22} & -\gamma_{12}\,\rho_{12} & -\gamma_{13}\,\rho_{13} & -\gamma_{14}\,\rho_{14} & -\gamma_{15}\,\rho_{15}\\
-\gamma_{21}\,\rho_{21} & \Gamma_3\,\rho_{33}-\Gamma_2\,\rho_{22} & -\gamma_{23}\,\rho_{23} & -\gamma_{24}\,\rho_{24} & -\gamma_{25}\,\rho_{25}\\
-\gamma_{31}\,\rho_{31} & -\gamma_{32}\,\rho_{32} & \Gamma_4\,\rho_{44}-\Gamma_3\,\rho_{33} & -\gamma_{34}\,\rho_{34} & -\gamma_{35}\,\rho_{35}\\
-\gamma_{41}\,\rho_{41} & -\gamma_{42}\,\rho_{42} & -\gamma_{43}\,\rho_{43} & \Gamma_5\,\rho_{55}-\Gamma_4\,\rho_{44} & -\gamma_{45}\,\rho_{45}\\
-\gamma_{51}\,\rho_{51} & -\gamma_{52}\,\rho_{52} & -\gamma_{53}\,\rho_{53} & -\gamma_{54}\,\rho_{54} & -\Gamma_5\,\rho_{55}
\end{pmatrix}
\label{L}
\end{equation}
\end{figure*}
The coherence between states \(|2\rangle\) and \(|1\rangle\) (\(\rho_{21}\)) obtained from the Lindblad master equation provides information about the probe transmission.
By substituting \(\boldsymbol{H} = \frac{\hbar}{2} \boldsymbol{M}\) into~\eqref{lind}, we obtain:

\begin{equation}
\frac{d\boldsymbol{\rho}}{dt} = -\frac{i}{2}\,(\boldsymbol{M}\boldsymbol{\rho} - \boldsymbol{\rho}\,\boldsymbol{M}) + \mathcal{L}(\boldsymbol{\rho}),
\end{equation}
or more specifically:
\begin{equation}
\frac{d\rho_{ij}}{dt}
= -\frac{i}{2}
\sum_{k=1}^{5} \left( M_{ik}\,\rho_{kj} - \rho_{ik}\,M_{kj} \right)
+ \mathcal{L}(\boldsymbol{\rho})_{ij}.
\label{Lind_new}
\end{equation}

Solving the Lindblad master equation for a five-level system requires handling 25 coupled differential equations, one for each density matrix element~$\boldsymbol{\rho}$.
 Several approximations are introduced to evaluate the effect of interference, analytically. However, to maintain precision in simulations, we used the QuTiP Python package~\cite{qutip5}, which numerically solves all the differential equations.

We assume a \textit{steady-state} condition, where the system has reached equilibrium and the probe signal is measured only after stabilization. This implies $\frac{d\rho_{21}}{dt} = \frac{d\rho_{31}}{dt} = \frac{d\rho_{41}}{dt} = \frac{d\rho_{51}}{dt} = 0$.

Next, the \textit{weak probe approximation} is applied: the system is assumed to remain mostly in the ground state, i.e., $\rho_{11} \approx 1$, while the excited-state populations are negligible, $\rho_{22}, \rho_{33}, \rho_{44}, \rho_{55} \ll 1$.
Finally, off-diagonal coherences such as $\rho_{32}, \rho_{42}, \rho_{52}$, which involve weakly populated and far-detuned states, are considered negligible due to the lack of strong direct coupling. 
With these approximations, the full set of 25 equations for solving $\rho_{21}$ reduces to the following:
{\small
\begin{subequations}
\begin{equation}
\left(-i\Delta_p + \gamma_{21}\right)\rho_{21}
+ \tfrac{i}{2}\,\Omega_c\,\rho_{31}
= -\tfrac{i}{2}\,\Omega_p
\end{equation}

\begin{equation}
\tfrac{i}{2}\,\Omega_{c}\,\rho_{21}
+ \left[-i(\Delta_{p}+\Delta_{c}) + \gamma_{31}\right]\,\rho_{31}
+ \tfrac{i}{2}\,\Omega_{\mathrm{RF}}\,\rho_{41}
= 0
\end{equation}

\begin{equation}
\tfrac{i}{2}\,\Omega_{\mathrm{RF}}\,\rho_{31}
+ \left[-i(\Delta_{p}+\Delta_{c}+\Delta_{\mathrm{RF}}) + \gamma_{41} \right]\,\rho_{41}
+ \tfrac{i}{2}\,\Omega_{I}\,\rho_{51}
= 0
\end{equation}

\begin{equation}
\tfrac{i}{2}\,\Omega_{I}\,\rho_{41}
+ \left[-i(\Delta_{p}+\Delta_{c}+\Delta_{\mathrm{RF}}+\Delta_{I}) + \gamma_{51} \right]\,\rho_{51}
= 0.
\end{equation}
\end{subequations}}

\begin{figure*}[t]
  \centering
  \footnotesize
  \setlength{\arraycolsep}{2pt}

\begin{equation}
\rho_{21}
= -\,\frac{\displaystyle \tfrac{i}{2}\,\Omega_p}
{%
\displaystyle
\bigl(-i\Delta_p+\gamma_{21}\bigr)
- \frac{\displaystyle \bigl(\tfrac{i}{2}\,\Omega_c\bigr)^{2}}
       {\displaystyle
        \bigl[-\,i(\Delta_p+\Delta_c)+\gamma_{31}\bigr]
        - \frac{\displaystyle \bigl(\tfrac{i}{2}\,\Omega_{\mathrm{RF}}\bigr)^{2}}
               {\displaystyle
                \bigl[-\,i(\Delta_p+\Delta_c+\Delta_{\mathrm{RF}})+\gamma_{41}\bigr]
                - \frac{\displaystyle \bigl(\tfrac{i}{2}\,\Omega_{I}\bigr)^{2}}
                       {\displaystyle
                        -\,i(\Delta_p+\Delta_c+\Delta_{\mathrm{RF}}+\Delta_{I})
                        +\gamma_{51}
                       }
               }
       }
}
\label{rho2121}
\end{equation}

\begin{equation}
\label{Stark}
\rho_{21}
\approx -\,\frac{\displaystyle \tfrac{i}{2}\,\Omega_p}
{%
\displaystyle
\left(-i\Delta_p + \gamma_{21}\right)
- \frac{\displaystyle \left(\tfrac{i}{2}\,\Omega_c\right)^2}
       {\displaystyle
        \left[-i(\Delta_p + \Delta_c) + \gamma_{31}\right]
        - \frac{\displaystyle \left(\tfrac{i}{2}\,\Omega_{\mathrm{RF}}\right)^2}
               {\displaystyle
                -i\left(\Delta_p + \Delta_c + \Delta_{\mathrm{RF}} - 
                \frac{\Omega_I^2}{4\Delta_I}
                \right)
                + \gamma_{41}
               }
       }
}
\end{equation}

  \label{fig:rho21}
\end{figure*}
After solving the system of linear equations, we can calculate $\rho_{21}$ as shown in~\eqref{rho2121} at the top of the next page. 
Since the interference signal is off-resonance, we have $\Delta_I \gg \Delta_p, \Delta_c, \Delta_{\mathrm{RF}}, \gamma_{51}$. 
Under this condition,~\eqref{rho2121} simplifies to~\eqref{Stark}, shown near the top of the next page. This equation shows that when off-resonant interference is present in a five-level system, it acts as an effective shift in the RF detuning. In other words, we have 
$\Delta_{\mathrm{RF}}' = \Delta_{\mathrm{RF}} - \frac{\Omega_I^2}{4\Delta_I}$. 
This means the probe laser readout will resemble that of a four-level system with RF signal radiation and without interference, but with an effective detuning of $\Delta_{\mathrm{RF}}'$.
This can be explained by the fact that off-resonant exposure cannot effectively couple the energy levels, but instead induces an AC Stark shift, i.e., a shift in the energy levels proportional to the intensity of the applied field~\cite{acstark}.

\begin{figure}
  \centering
  \includegraphics[width=0.45\textwidth]{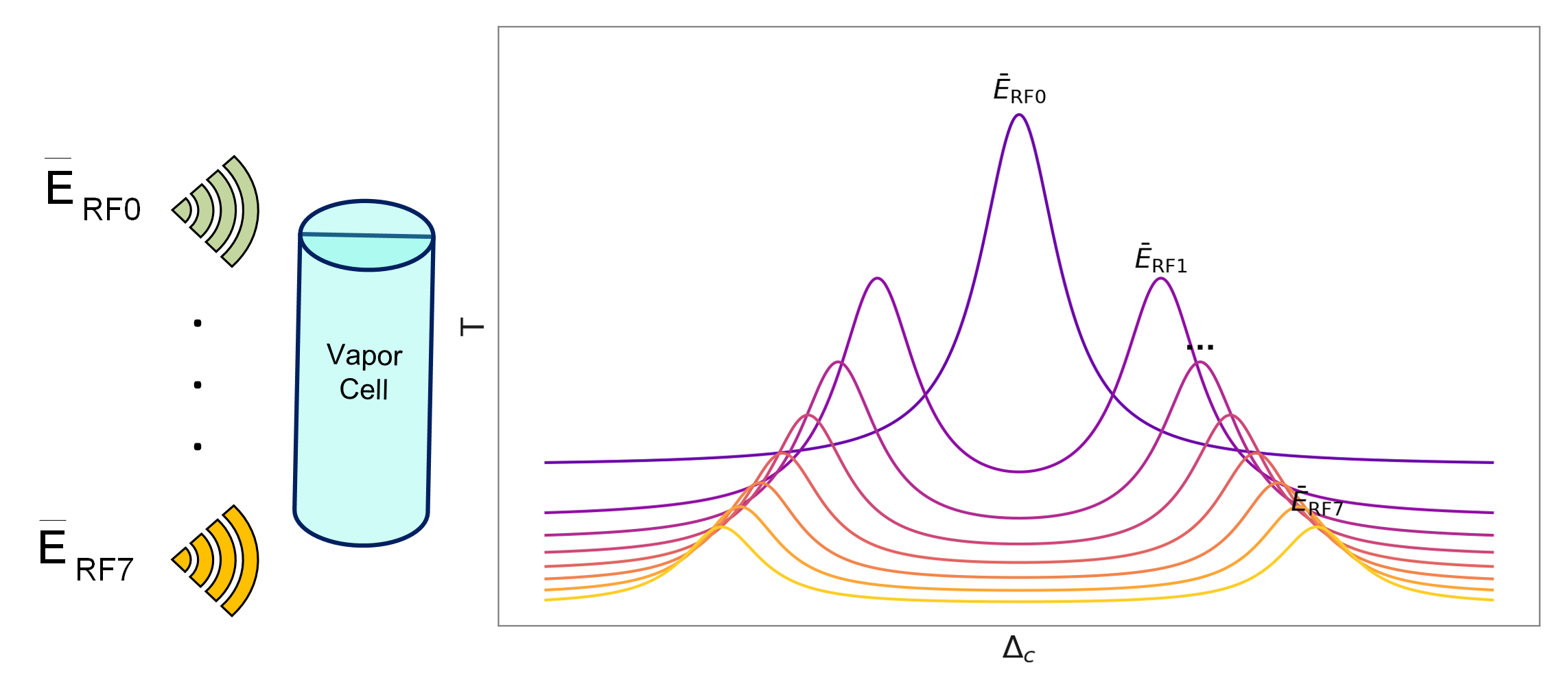}
  \caption{Probe Transmission of an 8-PAM receiver. }
  \label{8PAM}
\end{figure}

The goal is to calculate the transmission of the probe laser, which reflects the change in laser power after passing through the vapor cell. This can be described by the Beer–Lambert law~\cite{Beer1}:

\begin{equation}
T = \exp\left[
 \frac{4 \pi N L |\mu_{21}|^2}
       {\hbar \epsilon_0 \lambda_p \Omega_p}
\, \mathrm{Im}(\rho_{21})
\right],
\label{T}
\end{equation}
which shows that the probe transmission is proportional to $\mathrm{Im}(\rho_{21})$. Here, $N$ is the atomic density, $L$ is the length of the vapor cell, $\mu_{21}$ is the dipole moment of the transition $\lvert 2 \rangle \to \lvert 1 \rangle$, $\epsilon_0$ is the vacuum permittivity, and $\lambda_p$ is the wavelength of the probe laser.
From the behavior of the probe laser transmission, one can demodulate an 8-PAM-modulated signal, as shown in Fig.~\ref{8PAM}. This can be done in two ways. 

The first method, or frequency regime, measures peak-to-peak distance, proportional to the RF field amplitude \(\bar{E}_{\mathrm{RF}}\), according to the relation \(\Omega_{\mathrm{RF}} = \mu_{34} \bar{E}_{\mathrm{RF}} / \hbar\). When the RF signal is resonant, the peak-to-peak separation in the transmission spectrum is directly related to \(\Omega_{\mathrm{RF}}\). Therefore, a larger separation implies a stronger Rabi frequency and thus a higher RF field amplitude, enabling us to demodulate the PAM signal.

The second method, or amplitude regime, used here, reduces the need for wide frequency scanning. Instead of identifying peaks, we scan around the extremum between the two main peaks. The corresponding probe transmission value at that extremum represents different PAM levels. This approach significantly reduces the $\Delta_c$ scan points required, as the extrema occur within a fixed frequency interval for all signal amplitudes, whereas the peak positions shift with field intensity.

\section{Noise Model}
\label{noise}

It should be noted that $T$ in~\eqref{T} represents the ideal probe power transmission at the photodetector, and noise must be taken into account. There are two types of noise: intrinsic and extrinsic~\cite{noise}.

Background noise, such as ambient light is considered extrinsic. 
In this paper, we assume that with proper system design, extrinsic noise can be effectively mitigated and rendered negligible.
However, intrinsic noise must still be considered. It primarily consists of observation uncertainty and quantum projection noise. The former is modeled as Gaussian with zero mean and variance $\sigma^2_{\mathrm{UN}}$, arising from the photodetector’s limited precision. The uncertainty noise variance is signal dependent and can be modeled as proportional to the RF field amplitude, resulting in a noise variance of $\sigma^2_{\mathrm{UN}} = \epsilon\, \bar{E}_{\mathrm{RF}}^2$, where $\epsilon$ is a scaling factor typically less than 1\%~\cite{noise,noise2}.


To model the quantum projection noise, we use the dephasing time \( T_2 \) (which characterizes the timescale over which coherence between quantum states is lost), the integration time \( T_i \) (which is the time over which all measurements are integrated), and \( N_R \) as the number of excited Rydberg atoms. The minimum detectable electric field \( \bar{E}_{\mathrm{RF}}^{\mathrm{min}} \) in $[\mathrm{V/m}]$—defined as the \textit{root-mean-square (RMS)} electric field amplitude limited by quantum projection noise—can be expressed as~\cite{projection_noise, shotnoise}:

\begin{equation}
\bar{E}_{\mathrm{RF}}^{\mathrm{min}} = \frac{2\pi \hbar}{|\mu_{34}| \sqrt{N_R T_i T_2}},
\end{equation}
where the corresponding noise variance can be modeled as \(\sigma^2_{\mathrm{QPN}} = \left( \bar{E}_{\mathrm{RF}}^{\mathrm{min}} \right)^2\)~\cite{towards}.
The intrinsic noise, as it arises from independent sources, can be modeled as the sum of noise components represented by additive Gaussian variables with zero mean and total variance of \( \sigma^2 = \sigma^2_{\mathrm{UN}} + \sigma^2_{\mathrm{QPN}} \), due to the independence of noise sources.

The measured probe transmission at the photodetector, denoted by \( \tilde{T} \), is a noisy version of the ideal transmission \( T \), and using a first-order Taylor expansion can be approximated as

\begin{equation}
\tilde{T} = T + \frac{dT}{dE} \cdot n ,
\label{derivative}
\end{equation}
where \(n \sim \mathcal{N}(0, \sigma^2)\).
To compare the performance of a Rydberg receiver with a conventional receiver, we calculate SER, which quantifies the probability of incorrectly detecting a transmitted symbol. Conventional receivers use electronic circuitry and are subject to inevitable thermal noise. They also include filters designed to mitigate interference, with the degree of attenuation depending on the specific filter design.

We calculate the SER for the Rydberg receiver through simulation by evaluating how the noisy transmission signal $\tilde{T}$ deviates from the actual transmission $T$. A symbol error occurs when $\tilde{T}$ falls outside the correct decision region in the PAM demodulation constellation.

In a conventional system, SER in an \( M \)-PAM modulation scheme over an additive white Gaussian noise channel with noise spectral density \( N_0 \) and average symbol energy \( \mathcal{E}_s \) is given by~\cite{proakis2008digital}:
\begin{equation}
\mathrm{SER} = 2\left(1 - \frac{1}{M}\right) Q\left( \sqrt{ \frac{6 \mathcal{E}_s}{(M^2 - 1) N_0} } \right).
\label{SER}
\end{equation}

After the receiver filter, the effective noise power can be defined as \( N_{\text{eff}} = N_0 + N_{\text{I}} \), where \( N_{\text{I}} \) represents the power of the interference signal after filtering. The SER in our conventional receiver can be calculated using~\eqref{SER}, with \( N_0 \) replaced by the effective noise power \( N_{\text{eff}} \) to account for both thermal noise and residual interference.

\begin{figure*}[t]  
  \centering
  \begin{subfigure}[t]{0.41\textwidth}
    \centering
    \includegraphics[width=\textwidth]{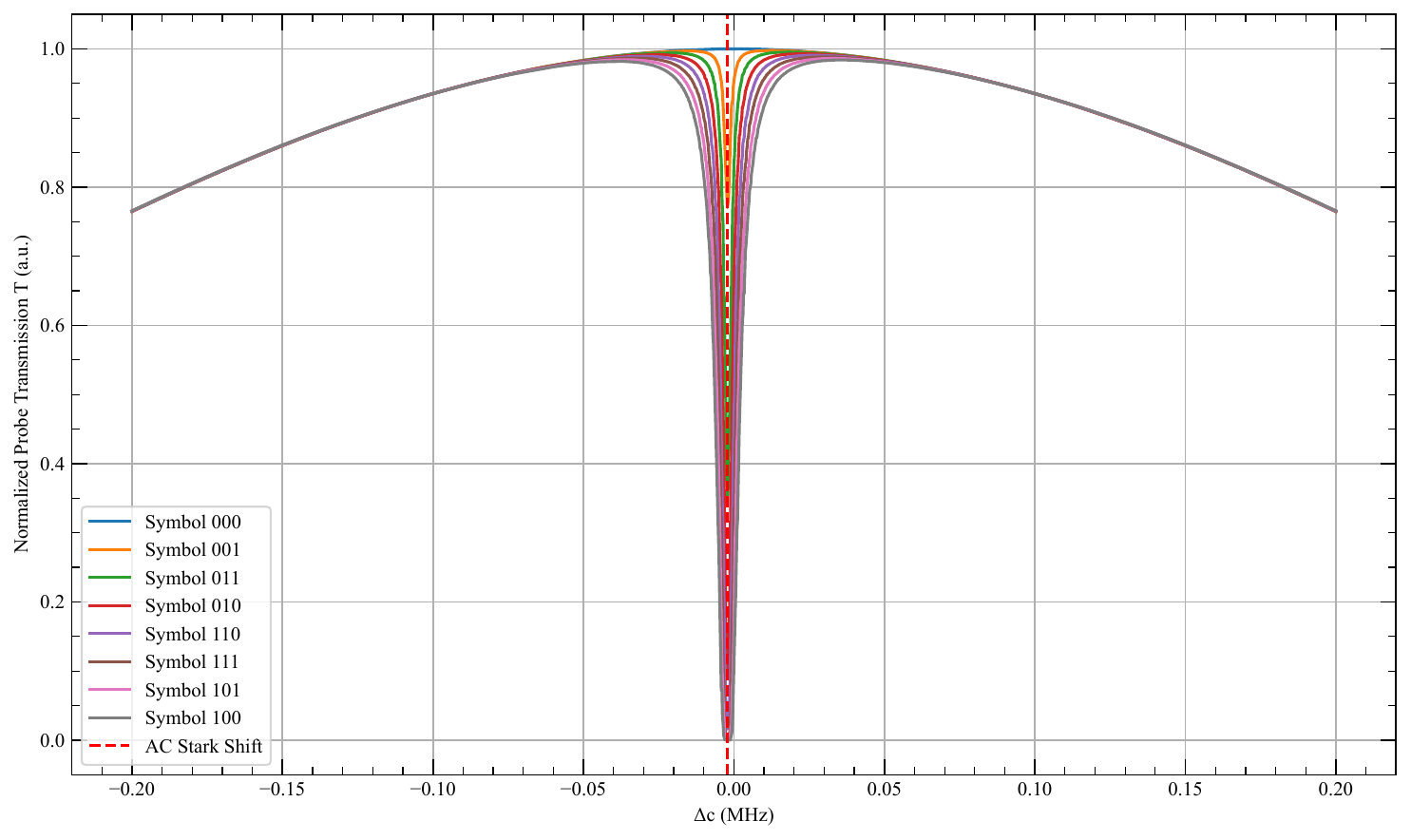}
    \caption{}
    \label{a}
  \end{subfigure}
  \hfill
  \begin{subfigure}[t]{0.41\textwidth}
    \centering
    \includegraphics[width=\textwidth]{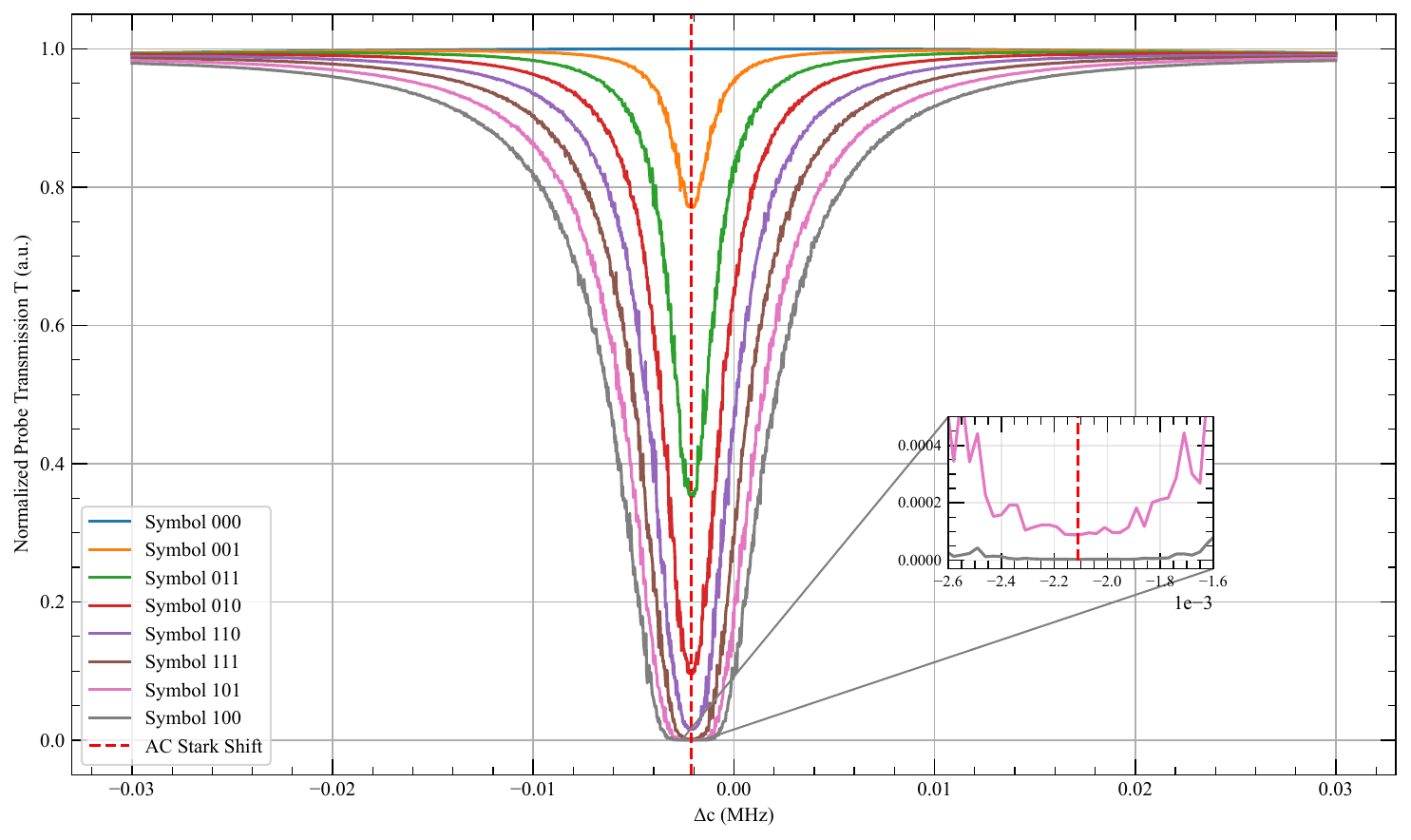}
    \caption{}
    \label{b}
  \end{subfigure}
  \caption{{Normalized probe transmission in an 8-PAM Rydberg receiver: (a) overall view, (b) zoomed-in view.
}}
  \label{sim1}
\end{figure*}
\section{Simulations}
\label{simulation}
The following parameters are used in the simulation. The decay rates are set as: $\Gamma_1 = 0$, $\Gamma_2 = 2\pi \times 6\,\mathrm{MHz}$, $\Gamma_3 = 2\pi \times 3\,\mathrm{kHz}$, and $\Gamma_4 = \Gamma_5 = 2\pi \times 2\,\mathrm{kHz}$~\cite{decay1,decay2}. As stated in~\cite{noise2}, the electric field of the RF signal inside the vapor cell can be measured with an uncertainty of \(\epsilon = 0.5\%\). The peak electric fields used in the simulation are \(1~\mathrm{V/m}\) for the probe field, \(80~\mathrm{kV/m}\) for the coupling field, and \(1~\mathrm{V/m}\) for the interference field. These values represent the maximum amplitudes of the time-varying electric fields applied to each transition. 

As the energy difference between states \(\lvert 4 \rangle\) and \(\lvert 5 \rangle\) corresponds to a frequency difference of \(34~\mathrm{GHz}\), and the interference field originates from a mid-band 5G signal, we approximate \(\Delta_I\) as \(-31~\mathrm{GHz}\). The full width at half maximum (FWHM) of the coupling and probe lasers is {\(100~\mathrm{Hz}\)}. This implies, assuming a Gaussian distribution for the laser beam, a standard deviation of approximately {\(42.4~\mathrm{Hz}\)}, i.e., \(\mathrm{FWHM}/\sqrt{8\ln 2}\)~\cite{FWHM}. Without loss of generality, the detunings are set as \(\Delta_p = 0\) and \(\Delta_{\mathrm{RF}} = 0\). The number of effective Rydberg atoms is \(N_R = 0.5 \times 10^{5}\) and the atomic density is \(N = 10^{17}~\mathrm{atoms/m^3}\); the vapor cell length is \(L = 75~\mathrm{mm}\), and the probe laser wavelength is \(\lambda_p = 780~\mathrm{nm}\). The dephasing time for the RF coupling can be calculated as \( T_2 = \frac{1}{0.5 (\Gamma_3 + \Gamma_4)} \), and the integration time is set to \( T_i = 100~\mu\mathrm{s} \). The electric field step size used in the derivative calculation in~\eqref{derivative} is \( \Delta E = 0.1~\mathrm{nV/cm} \). The 8-PAM signal levels are given by \( [0, 1, 2, 3, 4, 5, 6, 7] ~\mu\mathrm{V/cm} \).
{ Nonlinear level spacing could be explored in future work to increase the separation between higher levels.
}

The QuTiP package is used to solve the differential equations in~\eqref{Lind_new}, and the ARC library~\cite{arc2017} is employed to compute transition moments.
The probe laser and coupling field are applied through the vapor cell in opposite directions, and an RF 8-PAM modulated signal and an interference field are radiated through horn antennas toward the cell. We record the simulated normalized probe transmission after it passes through the vapor.

To demodulate the signal, we must distinguish between different symbols based on the probe transmission. Our approach relies on probe measurement in the amplitude regime, which focuses on the extremum at the center of the splitting—shifted from zero in our case due to the AC Stark effect. This method ideally requires only a single scan over the detuning frequency rather than sweeping the full spectrum. 

In practice, depending on the laser linewidth, additional measurements around \(\Delta_c\) equal to the AC Stark shift may be needed. However, in these simulations, we measure the probe transmission only at \(\Delta_c =\) AC Stark shift. As the interference intensity is unknown on the receiver side, the frequency for demodulation is also unknown. To find the extremum corresponding to the AC Stark shift, we calibrate the receiver by sending pilot signals and averaging the frequencies at which the minimum occurs. This calibration can introduce errors due to non-ideal laser stability. In the simulations, we also show the effect of this error on demodulator performance.


The performance of the conventional receiver is calculated using~\eqref{SER}, where the average symbol energy is computed as \(\mathcal{E}_s = \mathrm{mean}(\bar{E}_{\mathrm{RF}})^2 / (2 Z_0) \cdot T_s \cdot A_{\mathrm{eff}}\). Here, the factor of 2 accounts for the conversion from peak to RMS power, \(Z_0 = 377~\Omega\) is the impedance of free space, and \(T_s = T_i =100~\mu\mathrm{s}\) is the symbol duration. The effective antenna area is given by \(A_{\mathrm{eff}} = \lambda^2 G / (4\pi)\), where \(\lambda\) is the wavelength of the RF signal and \(G = 1.5\) for a conventional receiver. To estimate the energy from noise and interference, the interference energy is calculated in the same manner, assuming it passes through a filter with different attenuation levels.
Thermal noise is calculated as \(kT\), where \(k\) is the Boltzmann constant and \(T\) is the temperature, which we consider to be \(T = 290~\mathrm{K}\) (room temperature).

Fig.~\ref{a} shows the normalized probe transmission versus \(\Delta_c\), exhibiting ATS due to the presence of an RF signal at different electric field intensities, corresponding to the symbols of an 8-PAM modulator. The AC Stark shift appears as a shift in the \(\Delta_c\) position of the center of the splitting. Fig.~\ref{b} presents the same probe transmission, but zoomed in to highlight the distinct transmission levels associated with each PAM symbol, enabling symbol demodulation.

\begin{figure*}[t]  
  \centering
  \begin{subfigure}[t]{0.41\textwidth}
    \centering
    \includegraphics[width=\textwidth]{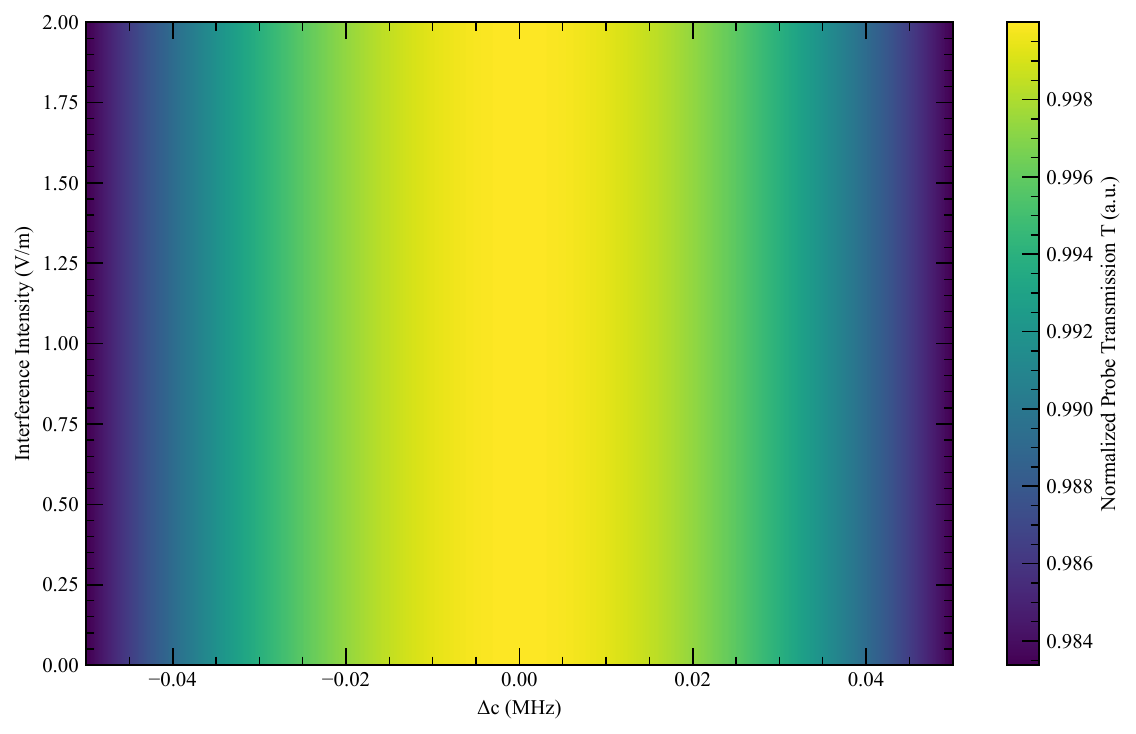}
    \caption{}
    \label{2a}
  \end{subfigure}
  \hfill
  \begin{subfigure}[t]{0.41\textwidth}
    \centering
    \includegraphics[width=\textwidth]{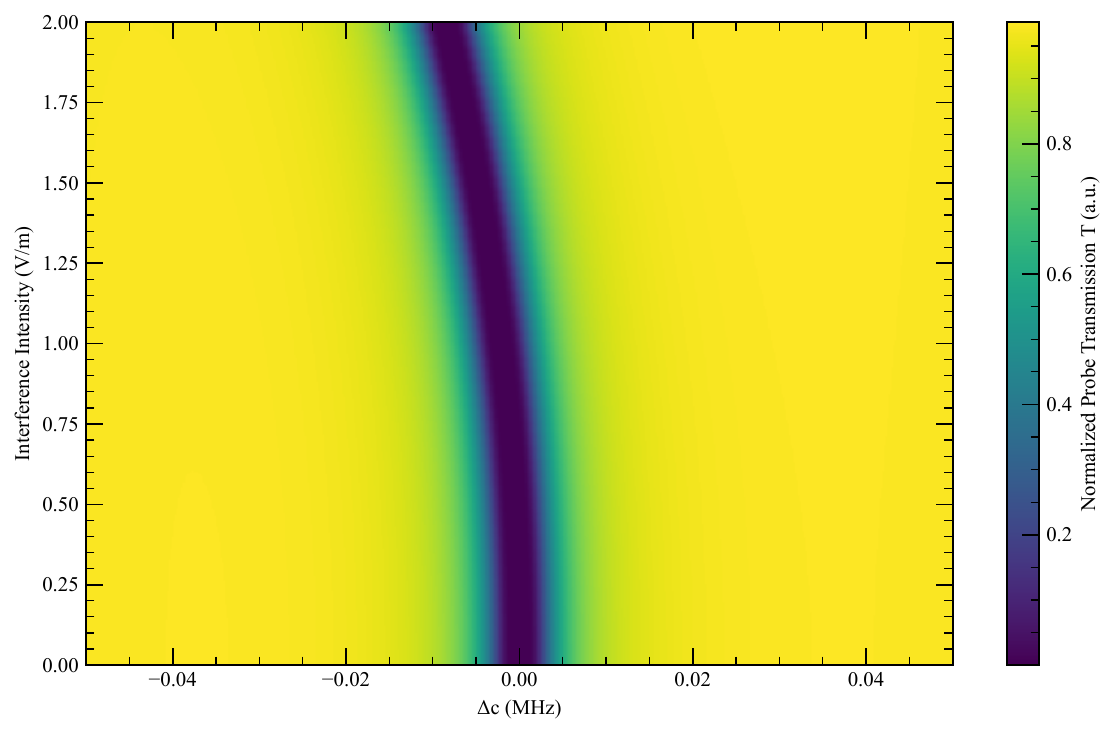}
    \caption{}
    \label{2b}
  \end{subfigure}
  \caption{Normalized probe transmission versus $\Delta_c$ and interference intensity: (a) without RF signal; (b) with RF signal of $7~\mu\mathrm{V/cm}$.}
  \label{sim2}
\end{figure*}

Fig.~\ref{2a} shows the normalized probe laser transmission versus interference intensity and \(\Delta_c\). As observed, the interference field has no effect on the transmission in the absence of an RF field. This behavior can be explained by~\eqref{rho2121}, where, when \(\Omega_{\mathrm{RF}} = 0\), the interference field alone cannot induce splitting and its effect is canceled within the corresponding fractional term. Fig.~\ref{2b} presents a similar plot, but with an RF field applied with a peak electric field of \(7~\mu\mathrm{V/cm}\). In this case, ATS becomes visible, and the AC Stark shift increases with the intensity of the interference field.

Table~\ref{table1} compares the SER of the conventional receiver and the Rydberg receiver under different design scenarios. The SER for the Rydberg receiver is numerically calculated by demodulating \(10^7\) symbols and summing up the errors. The conventional receiver uses filters with varying attenuation levels, while the Rydberg receiver operates with different calibration accuracies. {The accuracies are defined as the percentage deviation of the measurement 
at the AC Stark--shifted extremum, 
relative to the default case without interference (which occurs at 
$\Delta_c = 0$).}
It can be seen that even with a high calibration error, the Rydberg receiver still outperforms the conventional receiver equipped with an 85~dB interference attenuation filter, which represents a high-quality filter.

\begin{table}[htbp]
  \centering
  \small
  \caption{SER of conventional and Rydberg receivers under different parameters.}
  \label{table1}
  \begin{tabular}{cc}
    \toprule
    \textbf{Conventional Receiver} & \textbf{Rydberg Receiver} \\
    \midrule
    \begin{tabular}[t]{@{}ll@{}}
    \textbf{Filter Attenuation (dB)} & \textbf{SER} \\
    70 & \(5.9\mathrm{e}{-1}\) \\
    75 & \(4.1\mathrm{e}{-1}\) \\
    80 & \(1.7\mathrm{e}{-1}\) \\
    85 & \(1.9\mathrm{e}{-2}\) \\
    \end{tabular}
    &
    \begin{tabular}[t]{@{}ll@{}}
    \textbf{Accuracy (\%)} & \textbf{SER} \\
    40 & {\(3.9\mathrm{e}{-3}\)} \\
    60  & {\(8.3\mathrm{e}{-3}\)} \\
    80  & {\(4.3\mathrm{e}{-3}\)} \\
    100  & {\(1.2\mathrm{e}{-6}\)} \\
    \end{tabular}
    \\
    \bottomrule
  \end{tabular}
\end{table}
\section{Conclusions}
\label{conclusion}

In this paper, we investigated the use of a Rydberg receiver to mitigate interference and compared its performance with that of a conventional receiver. An 8-PAM modulated signal at 14.2~GHz, accompanied by interference in the mid-band 5G frequency range, was processed using a Rydberg receiver modeled as a five-level atomic system, where the interference was off-resonant. We demonstrated that the interference induces an AC Stark shift, which manifests as a shift in the probe laser transmission due to energy level perturbations. When properly calibrated, the Rydberg receiver can mitigate the effect of the interference and accurately demodulate the signal. Our results show that the SER of the Rydberg receiver outperforms that of a conventional receiver, even in the presence of calibration errors. {Future work will expand this study by validating the findings in an experimental setting.}

	\bibliography{Bibliography}
	\bibliographystyle{IEEEtran}

\end{document}